\documentclass[preprintnumbers,reprint,superscriptaddress,twocolumn]{revtex4}

\usepackage{graphicx}
\usepackage{dcolumn}
\usepackage{bm}
\usepackage{amssymb}
\usepackage{amsmath}
\usepackage{epsfig}    
\usepackage{color}
\usepackage{slashed}

\def\be{\begin{equation}}
\def\ee{\end{equation}}
\newcommand{\bea}{\begin{eqnarray}}
\newcommand{\eea}{\end{eqnarray}}

\numberwithin{equation}{section}

\begin{document} 
\title{Hidden Sector Dark Matter with 
Global $U(1)_X$-symmetry  and Fermi-LAT 130 GeV $\gamma$-ray Excess}
\preprint{KIAS-P13062}

\author{Seungwon Baek}
\email{swbaek@kias.re.kr}
\affiliation{School of Physics, KIAS, Seoul 130-722, Korea}
\author{Hiroshi Okada}
\email{ hokada@kias.re.kr}
\affiliation{School of Physics, KIAS, Seoul 130-722, Korea}



\begin{abstract}
We suggest a dark matter scenario which  could contribute the possible anomaly observed by Fermi-LAT 
$\gamma$-ray space telescope. It is based on the model recently proposed by Weinberg.
In our scenario the  gamma-ray line signal comes from the fermionic dark matter ($M_{\rm DM}=214 $ GeV) 
annihilating 
into two light scalars with mass around 500 MeV which in turn decay into two neutral pions. Finally the pions can
decay into two 130 GeV photons. The strong constraint from the direct detection
leaves only  the channel of the dark matter annihilation into two light scalars 
 for both the relic density and the Fermi-LAT  gamma-ray line signal.
The resulting gamma-ray spectrum is rather broad and does not fit to the data perfectly, but
the data also show there may be fluctuation in the spectrum. There is no associated $Z$-boson or Higgs boson
production contrary to most other works where the signal comes from the loops of charged particles.
The annihilation into the other SM particles are highly suppressed due to the small mixing from the
direct detection.
Future experiments with more data will give more clues on the possible scenarios.
\end{abstract}


%
\maketitle
\section{Introduction\label{sec:intro}}
The cosmological and astrophysical observations suggest that
27\% of the energy density of the universe is in the form
of dark matter (DM) \cite{Komatsu:2010fb, Ade:2013zuv}.
The most promising candidate for DM is so-called weakly interacting 
massive particle (WIMP). In that case we may study its nature through 
creation at accelerators such as LHC, the scattering with ordinary matter, 
or the pair annihilation into ordinary standard model (SM) particles
including photon~\cite{Zeldovich:1980st}.

The current DM density of the universe is related to the annihilation cross
section at the decoupling temperature as
\bea
 \Omega_{\rm DM} h^2 = \frac{3 \times 10^{-27} {\rm cm^3/s}}
{\langle \sigma v \rangle_{\rm th}}.\label{eq:cs-fermi}
\eea
Recently the analysis of FermiLAT gamma-ray data showed that there
may be some peak near 130 GeV, which can be interpreted as the
annihilation of DM~\cite{Bringmann:2012,Weniger:2012}.
This interpretation requires the annihilation cross section to be about 4\% of freeze-out cross
section:
\bea
 \langle \sigma v \rangle_{\gamma\gamma} = 0.042  \langle \sigma v \rangle_{\rm th}
=0.042 {\rm ~pb}\, c.
\label{eq:sigmav_Fermi}
\eea
Since we expect the annihilation into photons apparently come
from loop-induced process whose cross section is estimated to be
\bea
\frac{ \langle \sigma v \rangle_{\gamma\gamma}}{ \langle \sigma v \rangle_{\rm th}} = 
\left(\frac{\alpha_{\rm em}}{4\pi} \right)^2 \sim 10^{-7},
\eea
the observation calls for some non-conventional models.
There are many attempts  to explain the Fermi-LAT data with DM annihilation 
or decay by many authors \cite{recent,non-SUSY-1, non-SUSY-2, non-SUSY-3,Toma:2013bka, Giacchino:2013bta,BKS}. 
%
%
%

In this paper we introduce a new mechanism 
that  provides a possible explanation for the Fermi-LAT anomaly on the basis of  
a model recently proposed by Weinberg~\cite{Weinberg:2013kea}. 
This model is originally suggested in order to explain the possible deviation in  the effective neutrino 
number, $\Delta N_{\rm eff}=0.36 \pm 0.34$ at the 68~\% confidence level from
the Planck, WMAP9 polarization  and ground-based data~\cite{Ade:2013zuv},
although it is not very significant. We will just use the central value of the deviation
from now on.
 The author introduces a complex scalar field charged under a global $U(1)_X$ symmetry in the hidden sector.
All the SM particles are neutral under $U(1)_X$ and they interact with the hidden sector
via the renormalizable Higgs portal interaction~\cite{Higgs_portal}. 
The Goldstone boson (GB) after spontaneous symmetry breaking
can contribute to the relativistic energy density. Moreover he showed that a fermion in the hidden sector
can be introduced in such a way that it can carry odd parity after $U(1)_X$ is broken down to $Z_2$.
So the hidden fermion could be a promising DM candidate. 
Some authors analyze the Weinberg model in another aspects as well as LHC phenomenology ~\cite{Garcia-Cely:2013nin, Cheung:2013oya}.

To give the observed Fermi-LAT gamma-ray line signal, the DM is assumed to have mass around 214 GeV.
The correct thermal relic density is achieved by the DM interaction with the light scalar which decays dominantly
into the GB. It turns out that the same channel can accommodate the Fermi-LAT data when the light scalar decays
subdominantly into neutral pions.

This paper is organized as follows. In Section 2, we define our model based on Weinberg model. In Section 3, we discuss the possibility to explain  
the Fermi-LAT observation 
retaining consistency with  various other DM phenomenology. We conclude in Section 4.

\textcolor{red}{
}

\section{ Model}
In this section we set up a model, in which two new fields charged under global $U(1)_X$ symmetry $\chi$ and $\Psi_\pm$ are introduced
in addition to the SM fields. Here $\chi$ is boson with charge +2 
and $\Psi_\pm$ is fermion with charge +1 under the $U(1)_X$ symmetry. 
Notice that the SM fields are neutral under this symmetry. We expect the lighter one of $\Psi_\pm$ to be a DM candidate.

{\it Scalar sector}:
The new Lagrangian for the scalar sector is typically given by
\be
{\cal L}=
\partial_\mu\chi^*\partial^\mu\chi+\mu_\chi^2\chi^*\chi-\frac{\lambda_\chi}{2}(\chi^*\chi)^2-\lambda_{H\chi}(\Phi^\dag\Phi)(\chi^*\chi)
+{\cal L}_{\rm SM}, \label{lag-higgs}
\ee
where  $\mu_\chi^2$, $\lambda_\chi$, and $\lambda_{\chi\Phi}$ are real. Here we define the scalar fields in the unitary gauge  as follows:
\be
\Phi(x) =
\frac{1}{\sqrt{2}}
\left[
\begin{array}{c}
0\\
v_H+\phi(x)
\end{array}\right],\
\chi(x) =
\frac{1}{\sqrt{2}}
\left[
\begin{array}{c}
v_\chi+r(x)
\end{array}\right]e^{2i\alpha(x)},
\ee
where $ v_H^2=(246\ {\rm GeV})^2$ is  the vacuum expectation value (vev) of the SM, and $v_\chi$ is vev of the hidden sector, which can be determined
 by the analysis of DM data.
Then Eq. (\ref{lag-higgs}) can be rewritten as 
\bea
{\cal L}&=&
\frac12\partial_\mu r\partial^\mu r + 2(v_\chi+r)^2\partial_\mu\alpha\partial^\mu\alpha
+\frac12 \mu_\chi^2(v_\chi+r)^2
\nonumber\\
&-&\frac{\lambda_\chi}{8}(v_\chi+r)^4
-\frac12\lambda_{H\chi}(\Phi^\dag\Phi)(v_\chi+r)^2
+{\cal L}_{\rm SM}. \label{lag-higgs-expand}
\eea
The CP-even scalr mass-squared matrix in the basis of $(\phi,r)^t$ can be diagonalized by the following mixing matrix
\bea
 M^2_{\rm Higgs} &\equiv&
 \left(\begin{array}{cc} \lambda_Hv_H^2 & \lambda_{H\chi}v_Hv_\chi \\
 \lambda_{H\chi}v_Hv_\chi  & \lambda_\chi v_\chi^2 \end{array}\right)\\
\!\!&=&\!\!  \left(\begin{array}{cc} \cos\theta & \sin\theta \\ -\sin\theta & \cos\theta \end{array}\right)
\!\!\left(\begin{array}{cc} m^2_{{1}} & 0 \\ 0 & m^2_{{2}}  \end{array}\right)\!\!
\left(\begin{array}{cc} \cos\theta & -\sin\theta \\ \sin\theta & \cos\theta \end{array}\right),\nonumber
\label{eq:higgs}
\eea
where $\tan2\theta=2 \lambda_{H\chi}v_Hv_\chi/(\lambda_\chi v_\chi^2-\lambda_Hv_H^2)$.
The gauge eigenstate $(\phi,r)^t$ can be rewritten in terms of the mass eigenstate $(H_1,H_2)^t$
as
\begin{eqnarray}
 \phi &=& H_{1} \cos\theta + H_{2}\sin\theta, \nonumber\\
 r &=&- H_{1} \sin\theta + H_{2} \cos\theta.
\label{eq:mass_weak}
\end{eqnarray}
Hereafter we regard $H_1$ ($m_{1}$=125 GeV) as the SM Higgs boson, and $H_2$ as a lighter scalar boson 
whose mass is expected to be small of the order 500 MeV to accommodate a significant 
deviation in the effective neutrino number $\Delta N_{\rm eff}  =0.36$~\cite{Weinberg:2013kea}.

{\it Dark sector}:
The new Lagrangian for the DM sector is given by
\bea
{\cal L}&=& 
\frac{i}{2}\left(\bar\Psi_+\gamma^\mu\partial_\mu\Psi_+ + \bar\Psi_-\gamma^\mu\partial_\mu\Psi_-\right)\nonumber\\
&-&
\frac{i}{4v_\chi}\partial_\mu\alpha'\left(\bar\Psi_+\gamma^\mu\Psi_- - \bar\Psi_-\gamma^\mu\Psi_+\right)\nonumber\\
&-&
\frac{f}{2}(- H_{1} \sin\theta + H_{2} \cos\theta)\left(\bar\Psi_+\Psi_+ - \bar\Psi_-\Psi_-\right)\nonumber\\
&-&
\frac{1}{2}\left(m_+\bar\Psi_+\Psi_+ +m_- \bar\Psi_-\Psi_-\right),
\label{int-dm}
\eea
where we redefined $\alpha=\alpha'/(2v_\chi)$. Here we can take $f>0$ without loss of generality. 
Then $\Psi_-$ is a DM candidate with
mass $M_{\rm DM} = m_-$. We also obtain the mass difference, $\Delta m\equiv m_+ - m_- = 2 f v_\chi$.
It turns out that $\Delta m$ is very large in our scenario as we will see later. To get the large
mass difference we need  some degree of fine-tuning to get $M_{\rm DM}$ at electroweak scale.

\section{ Dark Matter }
In our DM analysis, we focus on explaining $\gamma$-ray excess at 130 GeV reported by the Fermi-LAT experiment. 
It is however worth mentioning the constraints from the other experiments before we go to the main part.

{\it Invisible decay of SM Higgs}:
The current experiment at LHC tells us that the invisible branching ratio of the SM Higgs (${\rm B_{inv}}$) is conservatively 
estimated to be less than 20\%~\cite{Belanger:2013kya}.
There are two invisible modes: $H_1\to 2\alpha'$ and $H_1\to 2 {\rm DM}$, and their decay rates
($\Gamma_{\rm inv}$) are given by  \cite{Garcia-Cely:2013nin}
\bea
&&\Gamma_{\rm inv}\equiv  \Gamma(H_1\to2\alpha')+\Gamma(H_1\to2 {\rm DM}),\\
&&\Gamma(H_1\to2\alpha')=\frac{m_{1}^3}{32\pi v_H^2}\sin^2\theta,\\
&&\Gamma(H_1\to2{\rm DM})=\frac{f^2 \sin^2\theta}{16\pi m_1^2}(m_1^2-4 M_{\rm DM})^{3/2}.
\eea
However, we consider $M_{\rm DM} > m_1/2$ and the latter mode is forbidden kinematically in our scenario.
One obtains the following relation~\cite{Garcia-Cely:2013nin}
\be
\Gamma_{\rm inv}< \frac{{\rm B_{inv}}\cos^2\theta}{1-{\rm B_{inv}}}\Gamma^{\rm SM}_{\rm Higgs},
\ee
where $\Gamma^{\rm SM}_{\rm Higgs}$ is the total decay width of the SM Higgs boson and estimated as $4.1\times 10^{-3}$ GeV at $m_{1}$=125 GeV. 
The upper bound on the invisible decay of Higgs restricts $\theta \lesssim 0.06$. This constraint is much
weaker than that from the direct detection of DM.

Another decay mode is $H_1 \to 2 H_2$ whose rate is given by
\be
\Gamma(H_1\to2H_2)\simeq \frac{m_{1}^3(v_H\cos\theta-v_\chi\sin\theta)^2}{128\pi v_H^2v_\chi^2}\sin^22\theta.
\label{eq:H22}
\ee
We assume this rate is about 10\% of that of $H_1 \to 2 \alpha^\prime$ to explain the
Fermi-LAT gamma-ray line.

{\it Direct detection}:
\begin{figure}[t]
\centering
\includegraphics[width=0.35\textwidth]{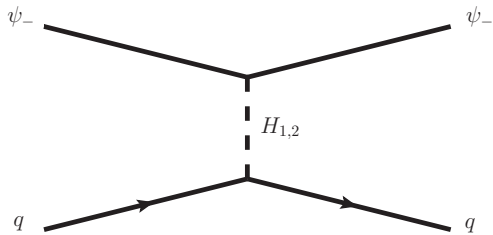}
\caption{The Feynman diagram for the elastic scattering of DM off the matter.
}
\label{fig:DD}
\end{figure}
The relevant process contributing to the spin independent scattering cross section is the $t-$channel
diagram mediated by   the lighter scalar as shown in Fig.~\ref{fig:DD}.
The corresponding elastic cross section is estimated as
\be
\sigma_p \approx 0.27^2 \frac{f^2 m_p^2 M_{\rm DM}^2}{4 \pi v_H^2 (M_{\rm DM} + m_p)^2 }
\left({1 \over m_1^2} -{1 \over m_2^2}\right)^2 \sin^2 2\theta,
\ee
where $m_p \approx 1$ GeV is the proton mass. 
It suggests the following constraint \cite{Garcia-Cely:2013nin}, which is derived from the current upper bound reported by XENON100 and 
LUX~\cite{xenon100}:
\be 
|f\sin2\theta| \le {\cal O}(10^{-5}).
\label{eq:DD}
\ee


{\it Fermi-LAT and Relic density}:
\begin{figure}[htb]
\centering
\includegraphics[width=0.15\textwidth,height=3cm]{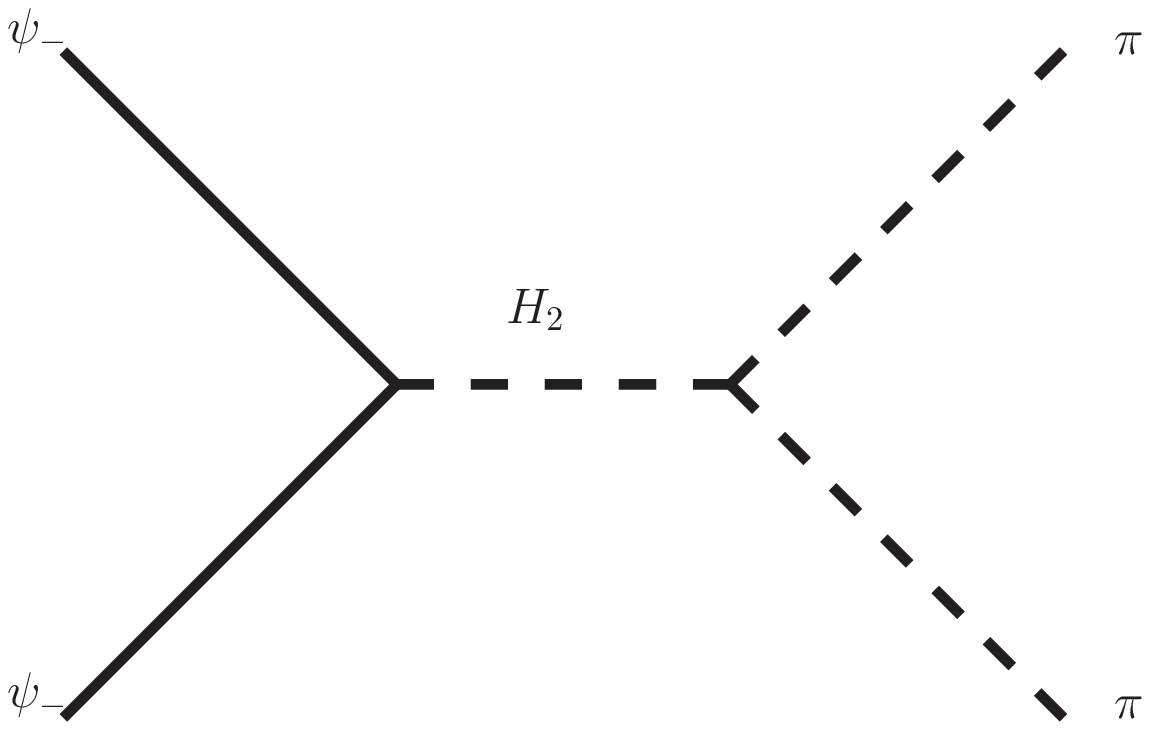}
\hspace{1cm}
\includegraphics[width=0.15\textwidth,height=3cm]{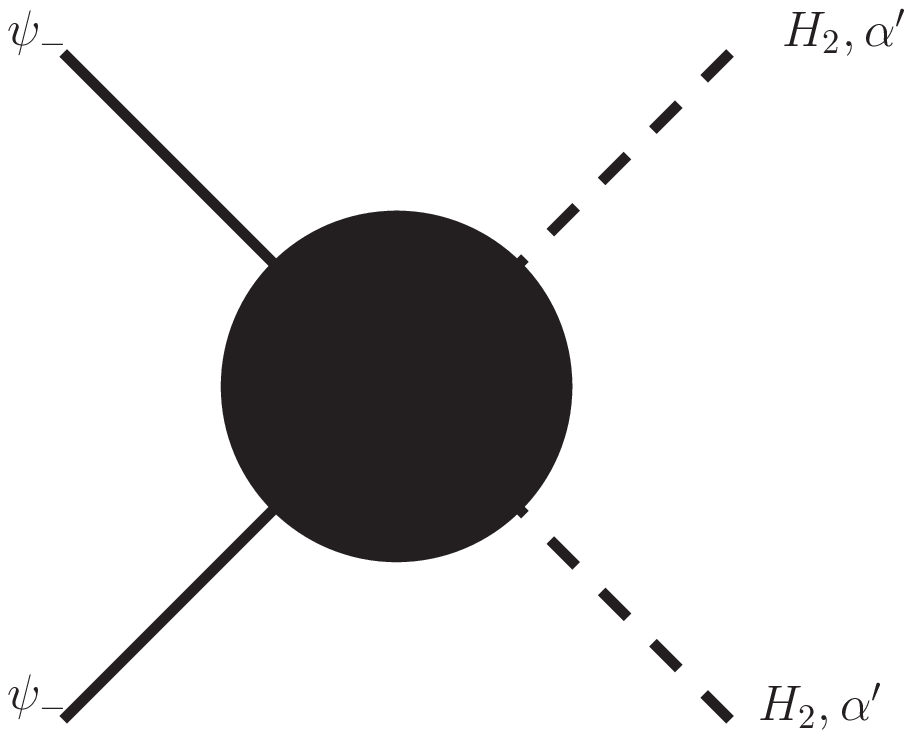}
\caption{The dominant Feynman diagrams for the relic density, where the black circle in the right panel represents all the possible channels at tree level.
}
\label{fig:ann}
\end{figure}

The possible dominant annihilation channels to obtain the current relic 
density are shown in Fig.~\ref{fig:ann}. They are i) 2 DM $\to 2 \pi$, ii) 2 DM $\to 2 H_2$,
iii) 2 DM $\to 2 \alpha^\prime$ and iv) DM coannihilation channels. But the channel i) is strongly
suppressed because its amplitude has the same parametric combination with Eq.~(\ref{eq:DD}). Therefore the
relic density is achieved either by one of ii), iii), iv) or combination of them. The Ref.~\cite{Garcia-Cely:2013nin}
shows that iv) is dominant when it is allowed, which is not allowed in our case.
The annihilation cross sections for the processes ii)  and iii) are estimated to be
\bea
\langle \sigma v \rangle_{2 H_2} \approx \frac{3 f^4 v_{\rm rel}^2}{128 \pi M_{\rm DM}^2}, 
\quad
\langle \sigma v \rangle_{2 \alpha^\prime} \approx \frac{ f^4 v_{\rm rel}^2}{32 \pi m_{+}^2}.
\eea
In our case it turns out the mode ii) is dominant because $M_{\rm DM} \ll m_{+}$.

The photon line observed by Fermi-LAT comes from ii) when $H_2 \to 2 \pi$.
 Then the pion decays into two photons. It implies that the mass  of DM is  fixed to be 
214 GeV~\footnote{If the gamma-ray line were emitted from $H_2$, its maximum energy would
be just $M_{\rm DM}$ In our case the peak energy $E_\gamma=130$ GeV is obtained
for $M_{\rm DM} = 214$ GeV, since it comes from the decay of $H_2$ into two pions.}.
For either case, the condition Eq.~(\ref{eq:sigmav_Fermi}) can be satisfied if
\bea
 {\rm B}(H_2 \to 2 \pi^0)\approx  1.1~\%,
 \label{eq:R}
\eea
because ${\rm B}(\pi^0\to2\gamma)\simeq 99\%$~\cite{Beringer:1900zz}.
This can be understood as follows: if we set $x_0 = {\rm B}(H_2 \to \pi^0 \pi^0)$,
$x_+ = {\rm B}(H_2 \to \pi^+ \pi^-)$, $x_\alpha = {\rm B}(H_2 \to \alpha' \alpha')$,
and considering $\psi_- \psi_- \to H_2 H_2$ dominates,
Eq.~(\ref{eq:sigmav_Fermi}) requires
\bea
4 x_0^2 + 2 (2 x_\alpha x_0) + 2 (2 x_0 x_+) \approx 4.2~\%.
\eea
Using $x_0 + x_+ + x_\alpha \approx 1$, we get $x_0 \approx 1.1~\%$.
We define ratio $R$ as
\bea
R \equiv \frac{ {\rm B}(H_2 \to 2 \pi^0)}{ {\rm B}(H_2 \to \alpha' \alpha')} \approx 1.1~\%.
\eea
The ratio $R$ is given in the Ref.~\cite{Cheung:2013oya},
\bea
R &=& \theta^2 \frac{v_\chi^2}{v_H^2} \left(1-\frac{4 m_\pi^2}{m_2^2}\right)^{1/2}
\left(1+\frac{2 m_\pi^2}{m_2^2}\right)^{2}\nonumber\\
& \approx& 0.011 \left(\theta \over 10^{-5} \right)^2
\left(v_\chi \over 2.5 \times 10^6 {\rm GeV} \right)^2.
\eea
We can see $f \approx 0.9$, $\theta \approx 10^{-5}$, and $v_\chi \approx  2.5 \times 10^6$ GeV
satisfies both Eq.~(\ref{eq:DD}) and Eq.~(\ref{eq:R}), using micromegas \cite{Belanger:2008sj}.
As a result we can obtain the correct
annihilation cross section necessary to explain the Fermi-LAT gamma-ray line.
It is worth mentioning that the shape line at 130 GeV is rather wide, if the photons are produced via 
neutral pions \cite{ non-SUSY-2, pion}, although the fall-off of the peak can be explained
and the data still show fluctuation in the spectrum. We also note that in our scenario
there is no associated $Z$-boson or Higgs boson
production contrary to most other works where the signal comes from the loops of charged particles.
Future experiments with more data will give more clues on the possible scenarios.

\section{Conclusions}
\label{sec:Conclusions}
We considered a dark global $U(1)_X$ model with a Goldstone boson and a dark matter~\cite{Weinberg:2013kea}.
The Goldstone boson can contribute to the effective neutrino number $\Delta N_{\rm eff} =0.36$
if the dark scalar mass is about 500 MeV. 
We showed that this light dark scalar produced by
the dark matter annihilation can mix with the SM Higgs boson and about 4\% of them 
can decay into two neutral pions. These pions finally decay into two photons with energy 130 GeV
if the dark matter mass is 214 GeV. Our benchmark parameters for the Fermi-LAT gamma-ray line are dark scalar coupling with the dark 
matter $\sim 1$, the mixing angle of the dark scalar with the SM Higgs $\sim 10^{-5}$, and
the vev of the dark scalar $\sim 3.5 \ \times 10^{6}$ GeV.

The dark matter relic density can possibly be obtained by four channels:
i) 2 DM $\to 2 \pi$, ii) 2 DM $\to 2 H_2$,
iii) 2 DM $\to 2 \alpha^\prime$ and iv) DM coannihilation.
But the strong constraint from the direct detection makes the channel i) always negligible.
The parameter space explaining the Fermi-LAT gamma-ray line makes the iii) and iv) suppressed.
Therefore only ii) is dominant contribution to the cross section for the relic density.

The obtained gamma-ray spectrum is broad box shape and does not fit to the data perfectly, but
the data show there may be fluctuation in the spectrum. There is no associated $Z\gamma$ or
$h \gamma$ production signal
 contrary to most other works where the signal comes from the loops of charged particles.
The annihilation into the other SM particles are highly suppressed due to the small mixing from the
direct detection, so we can avoid the constraints from the indirect detection easily.
Future experiments with more data will give more clues on the possible scenarios.

The generic signature of the model at the collider is  the production of $H_1$ via $g g \to H_1$ and its subsequent
decay $H_1 \to H_2 H_2 \to (\pi\pi) (\alpha^\prime \alpha^\prime)$~\cite{Cheung:2013oya}.
However, the branching ratio $\Gamma(H_1 \to 2 H_2)$ is very small in our scenario
as can be seen in Eq.~(\ref{eq:H22}).

\acknowledgments
We are grateful to P. Ko and T. Toma for fruitful discussions and KIAS Workshop in Jeju island for
providing us with the nice environment for discussion.
This work is partly supported by NRF Research Grant  2012R1A2A1A01006053 (SB).

\appendix


\begin{thebibliography}{999}


\bibitem{Komatsu:2010fb} 
  E.~Komatsu {\it et al.}  [WMAP Collaboration],
  Astrophys.\ J.\ Suppl.\  {\bf 192}, 18 (2011)
  [arXiv:1001.4538 [astro-ph.CO]].


\bibitem{Ade:2013zuv} 
  P.~A.~R.~Ade {\it et al.}  [Planck Collaboration],
  arXiv:1303.5076 [astro-ph.CO];
  C.~L.~Bennett {\it et al.}  [WMAP Collaboration],
  Astrophys.\ J.\ Suppl.\  {\bf 208}, 20 (2013)
  [arXiv:1212.5225 [astro-ph.CO]];
  S.~Das, T.~Louis, M.~R.~Nolta, G.~E.~Addison, E.~S.~Battistelli, J R.~Bond, E.~Calabrese and D.~C.~M.~J.~Devlin {\it et al.},
  arXiv:1301.1037 [astro-ph.CO];
  C.~L.~Reichardt, L.~Shaw, O.~Zahn, K.~A.~Aird, B.~A.~Benson, L.~E.~Bleem, J.~E.~Carlstrom and C.~L.~Chang {\it et al.},
  Astrophys.\ J.\  {\bf 755}, 70 (2012)
  [arXiv:1111.0932 [astro-ph.CO]].

\bibitem{Jarosik:2010}
  N.~Jarosik, C.~L.~Bennett, J.~Dunkley, B.~Gold, M.~R.~Greason, M.~Halpern, R.~S.~Hill, G.~Hinshaw {\it Eu AB.},
  Astrophys.\ J.\ Suppl.\  {\bf 192}, 14 (2011).
  [arXiv:1001.4744 [astro-ph.CO]].

\bibitem{Zeldovich:1980st} 
  Y.~.B.~Zeldovich, A.~A.~Klypin, M.~Y.~.Khlopov and V.~M.~Chechetkin,
  Sov.\ J.\ Nucl.\ Phys.\  {\bf 31}, 664 (1980)
  [Yad.\ Fiz.\  {\bf 31}, 1286 (1980)].

\bibitem{Bringmann:2012} 
  T.~Bringmann, X.~Huang, A.~Ibarra, S.~Vogl and C.~Weniger,
  arXiv:1203.1312 [hep-ph].

\bibitem{Weniger:2012} 
  C.~Weniger,
  arXiv:1204.2797 [hep-ph].




\bibitem{recent}
  T.~Basak and T.~Mondal,
  arXiv:1308.0023 [hep-ph];
  K.~-Y.~Choi and O.~Seto,
  Phys.\ Rev.\ D {\bf 86}, 043515 (2012)
  [Erratum-ibid.\ D {\bf 86}, 089904 (2012)]
  [arXiv:1205.3276 [hep-ph]];
  B.~Kyae and J.~-C.~Park,
  Phys.\ Lett.\ B {\bf 718}, 1425 (2013)
  [arXiv:1205.4151 [hep-ph]];
  D.~Das, U.~Ellwanger and P.~Mitropoulos,
  JCAP {\bf 1208}, 003 (2012)
  [arXiv:1206.2639 [hep-ph]];
  Z.~Kang, T.~Li, J.~Li and Y.~Liu,
  arXiv:1206.2863 [hep-ph];
  W.~Buchmuller and M.~Garny,
  JCAP {\bf 1208}, 035 (2012)
  [arXiv:1206.7056 [hep-ph]];
  J.~-C.~Park and S.~C.~Park,
  Phys.\ Lett.\ B {\bf 718}, 1401 (2013)
  [arXiv:1207.4981 [hep-ph]];
  J.~M.~Cline, A.~R.~Frey and G.~D.~Moore,
  Phys.\ Rev.\ D {\bf 86}, 115013 (2012)
  [arXiv:1208.2685 [hep-ph]];
  L.~Wang and X.~-F.~Han,
  Phys.\ Rev.\ D {\bf 87}, 015015 (2013)
  [arXiv:1209.0376 [hep-ph]];
  K.~Schmidt-Hoberg, F.~Staub and M.~W.~Winkler,
  JHEP {\bf 1301}, 124 (2013)
  [arXiv:1211.2835 [hep-ph]];
  Y.~Farzan and A.~R.~Akbarieh,
  arXiv:1211.4685 [hep-ph];
  G.~Chalons, M.~J.~Dolan and C.~McCabe,
  JCAP {\bf 1302}, 016 (2013)
  [arXiv:1211.5154 [hep-ph]];
  M.~Asano, T.~Bringmann, G.~Sigl and M.~Vollmann,
  arXiv:1211.6739 [hep-ph];
  C.~B.~Jackson, G.~Servant, G.~Shaughnessy, T.~M.~P.~Tait and M.~Taoso,
  arXiv:1302.1802 [hep-ph];
  J.~Kumar and P.~Sandick,
  arXiv:1303.2384 [hep-ph];
  A.~Ibarra, H.~M.~Lee, S.~L\'op.~Gehler, W.~-I.~Park and M.~Pato,
  arXiv:1303.6632 [hep-ph];
  K.~-Y.~Choi, H.~M.~Lee and O.~Seto,
  arXiv:1304.0966 [hep-ph];
E.~Dudas, Y.~Mambrini, S.~Pokorski and A.~Romagnoni,
  arXiv:1205.1520 [hep-ph];
H.~M.~Lee, M.~Park and W.~-I.~Park,
  arXiv:1205.4675 [hep-ph];
B.~S.~Acharya, G.~Kane, P.~Kumar, R.~Lu and B.~Zheng,
  arXiv:1205.5789 [hep-ph];
X.~Chu, T.~Hambye, T.~Scarna and M.~H.~G.~Tytgat,
  arXiv:1206.2279 [hep-ph];
N.~Weiner and I.~Yavin,
  arXiv:1206.2910 [hep-ph];
L.~Feng, Q.~Yuan and Y.~-Z.~Fan,
  arXiv:1206.4758 [astro-ph.HE];
  T.~Cohen, M.~Lisanti, T.~R.~Slatyer and J.~G.~Wacker,
  JHEP {\bf 1210}, 134 (2012)
  [arXiv:1207.0800 [hep-ph]];
J.~H.~Heo and C.~S.~Kim,
  arXiv:1207.1341 [astro-ph.HE];
E.~Hardy, J.~March-Russell and J.~Unwin,
  arXiv:1207.1435 [hep-ph];
M.~T.~Frandsen, U.~Haisch, F.~Kahlhoefer, P.~Mertsch and K.~Schmidt-Hoberg,
  arXiv:1207.3971 [hep-ph];
T.~Li, J.~A.~Maxin, D.~V.~Nanopoulos and J.~W.~Walker,
  arXiv:1208.1999 [hep-ph];
Y.~Bai and J.~Shelton,
  arXiv:1208.4100 [hep-ph];
 R.~Laha, K.~C.~Y.~Ng, B.~Dasgupta and S.~Horiuchi,
  arXiv:1208.5488 [astro-ph.CO];
L.~Bergstrom,
  arXiv:1208.6082 [hep-ph];
J.~Fan and M.~Reece,
  Phys.\ Rev.\ D {\bf 88}, 035014 (2013)
  [arXiv:1209.1097 [hep-ph]];
  N.~Bernal, C.~Boehm, S.~Palomares-Ruiz, J.~Silk and T.~Toma,
  Phys.\ Lett.\ B {\bf 723}, 100 (2013)
  [arXiv:1211.2639 [hep-ph]];
  M.~M.~Ettefaghi and R.~Moazzemi,
  JCAP {\bf 1302}, 048 (2013)
  [arXiv:1301.4892 [hep-ph]];
  L.~A.~Anchordoqui and B.~J.~Vlcek,
  Phys.\  Rev.\ D {\bf 88}, 043513 (2013)
  [Phys.\ Rev.\ D {\bf 88}, 043513 (2013)]
  [arXiv:1305.4625 [hep-ph]].
  

\bibitem{non-SUSY-1}
J.~M.~Cline,
  Phys.\ Rev.\ D {\bf 86}, 015016 (2012)
  [arXiv:1205.2688 [hep-ph]];
  \bibitem{non-SUSY-2}
  M.~R.~Buckley and D.~Hooper,
  Phys.\ Rev.\ D {\bf 86}, 043524 (2012)
  [arXiv:1205.6811 [hep-ph]];
  \bibitem{non-SUSY-3}
S.~Tulin, H.~-B.~Yu and K.~M.~Zurek,
  Phys.\ Rev.\ D {\bf 87}, 036011 (2013)
  [arXiv:1208.0009 [hep-ph]].
  
\bibitem{BKS}
S.~Baek, P.~Ko and E.~Senaha,
  arXiv:1209.1685 [hep-ph].



\bibitem{Toma:2013bka} 
  T.~Toma,
  Phys.\ Rev.\ Lett.\  {\bf 111}, 091301 (2013)
  [arXiv:1307.6181 [hep-ph]].

\bibitem{Giacchino:2013bta} 
  F.~Giacchino, L.~Lopez-Honorez and M.~H.~G.~Tytgat,
  arXiv:1307.6480 [hep-ph].

%
%
%

\bibitem{Weinberg:2013kea} 
  S.~Weinberg,
  Phys.\ Rev.\ Lett.\  {\bf 110}, 241301 (2013)
  [arXiv:1305.1971 [astro-ph.CO]].

\bibitem{Higgs_portal} 
  S.~Baek, P.~Ko and W.~-I.~Park,
  JHEP {\bf 1202}, 047 (2012)
  [arXiv:1112.1847 [hep-ph]];
S.~Baek, P.~Ko, W.~-I.~Park and E.~Senaha,
  JHEP {\bf 1211}, 116 (2012)
  [arXiv:1209.4163 [hep-ph]];

\bibitem{Garcia-Cely:2013nin} 
  C.~Garcia-Cely, A.~Ibarra and E.~Molinaro,
  arXiv:1310.6256 [hep-ph].

\bibitem{Cheung:2013oya} 
  K.~Cheung, W.~-Y.~Keung and T.~-C.~Yuan,
  arXiv:1308.4235 [hep-ph].



\bibitem{Beringer:1900zz} 
  J.~Beringer {\it et al.}  [Particle Data Group Collaboration],
  Phys.\ Rev.\ D {\bf 86}, 010001 (2012).

\bibitem{Belanger:2013kya} 
  G.~Belanger, B.~Dumont, U.~Ellwanger, J.~F.~Gunion and S.~Kraml,
  Phys.\ Lett.\ B {\bf 723}, 340 (2013)
  [arXiv:1302.5694 [hep-ph]].
  

 


\bibitem{xenon100} 
  E.~Aprile {\it et al.}  [XENON100 Collaboration],
  Phys.\ Rev.\ Lett.\  {\bf 109}, 181301 (2012)
  [arXiv:1207.5988 [astro-ph.CO]];
D.~S.~Akerib {\it et al.}  [LUX Collaboration],
  arXiv:1310.8214 [astro-ph.CO].


\bibitem{Belanger:2008sj} 
  G.~Belanger, F.~Boudjema, A.~Pukhov and A.~Semenov,
  Comput.\ Phys.\ Commun.\  {\bf 180}, 747 (2009)
  [arXiv:0803.2360 [hep-ph]].

\bibitem{pion} 
  E.~Tempel, A.~Hektor and M.~Raidal,
  JCAP {\bf 1209}, 032 (2012)
  [Addendum-ibid.\  {\bf 1211}, A01 (2012)]
  [arXiv:1205.1045 [hep-ph]];
  







\end{thebibliography}
\end{document}